\documentstyle{article}
\begin{document}
\begin{sloppypar}
\begin{center}

{\large \bf ON POSSIBILITY OF
HIGGS MECHANISM BREAK DOWN DUE TO THE INSTANTONS }

\vspace{40pt}
{\large Roman G. Shulyakovsky }\\

\vspace{30pt} {\it Institute of Physics, National Academy of
Sciences of Belarus, \\ F.Scorina av.,68, Minsk 220072\\ BELARUS
\\ e-mail: shul@dragon.bas-net.by}

\vspace{40pt}

ABSTRACT
\end{center}

\noindent 
Two-dimensional Abelian Higgs model is considered. It is shown
that Higgs mechanism of the gauge bosons masses generating is
broken in presence of instantons. It is supposed that discussion
can be generalized into the standard model of electroweak
interactions.

\section{Introduction}
It is well known that some field theories (including QCD and
Weinberg-Salam model) have the degenerated vacuum structure on the
classical level~\cite{JR}. Unequivalent classical vacua are
characterized by different Chern-Simons numbers $N$ and separated
by the energy barriers. Amplitude of the tunneling through these
barriers is nonzero if the theory admits
instantons~\cite{BPST,Hooft} (classical finite-action Euclidean
solutions). Thus classical degeneration disappears due to the
instantons. The correct vacuum state is approximately linear
combination of the naive classical vacua with different $N$.

Ordinary perturbative theory corresponds to the $N=0$. It is quite naturally
that taking into account of the another vacua with $N\neq 0$
leads to the essential consequences. Let us remind the most important
of them.

1. Instantons lead to the charge confinement in the 2-dimensional
Abelian Higgs model~\cite{CDG} and 3-dimensional
non-Abelian Higgs model~\cite{Pol}.

2.In the electroweak theory instantons induce baryon and lepton numbers
violation~\cite{Hooft} which can be connected with the matter and
antimatter asymmetry of the Universe~\cite{Zuccero}.

3.In QCD instantons lead to the quark and gluon condensates formation,
spontaneously break chiral symmetry, solve $U(1)$-problem~\cite{Hooft}.
It was suggested that QCD-instantons can be produced in deep-inelastic
scattering~\cite{Bal} and identified by means of the analysis of the
final states at HERA(DESY)~\cite{all}.

We would like to attract an attention to the once more consequence
of the instantons. It is known that in the Weinberg-Salam
theory~\cite{Weinberg} $W^{\pm}$ and $Z^0$ bosons get the masses
due to the interaction with Higgs fields, which have nonzero
vacuum expectation value and spontaneously break local $SU(2)$
symmetry~\cite{Higgs}. However Higgs mechanism of the masses
generating is perturbative phenomenon and it is violated, for
example, at a temperature higher than about $10^3\
GeV$~\cite{Linde} what could occur at the early Universe. At high
temperature local $SU(2)$ symmetry is restored and $W^{\pm}$ and
$Z^0$ bosons become massless. In this letter we demonstrate that
Higgs mechanism is broken down even at {\it zero temperature} if
we take into account instanton tunneling transitions. In sections
2 and 3 the example of 2-dimensional Abelian Higgs model is
considered. Section 4 is devoted to the possibility of Higgs
mechanism violation in Weinberg-salam theory.

\section{Local gauge symmetry restoration owing to the existence
of the instantons}
Abelian Higgs model describes the interactions of the real vector
fields $A_{\mu}(t,x)$ and self-interacting complex scalar fields $\phi (t,x)$:

\begin{equation}
L=-\frac{1}{4}F_{\mu\nu}F_{\mu\nu}+\frac{1}{2}(D_{\mu}\phi)^*D_{\mu}\phi
-\lambda(\phi^*\phi-\rho^2)^2,
\label{L}
\end{equation}
$$
D_{\mu}=\partial_{\mu}-ieA_{\mu},\quad
F_{\mu\nu}=\partial_{\mu}A_{\nu}-\partial_{\nu}A_{\mu}, \quad
\mu ,\nu =0,1;\quad \lambda >0,\ \rho >0.
$$

\noindent
There are the following classical vacua:

\begin{equation}
A_{\mu}(t,x)=\frac{1}{e}\partial_{\mu}\alpha (t,x),\quad
\phi (t,x)=e^{i\alpha (t,x)}\rho.
\label{vac}
\end{equation}

\noindent
In the perturbative approach vacuum $A_{\mu}(t,x)=0,\ \phi (t,x)=\rho$
is chosen.
Lagrangian for the small
perturbations near this vacuum

$$
L=-\frac{1}{4}F_{\mu\nu}F_{\mu\nu}+\frac{e^2\rho^2}{2}A_{\mu}A_{\mu}+
\frac{1}{2}e^2A_{\mu}A_{\mu}(2\eta\rho +\eta^2)+
$$
\begin{equation}
+\frac{1}{2}\partial_{\mu}\eta\partial_{\mu}\eta -
4\lambda\eta^2\rho^2-4\lambda\eta^3\rho-\lambda\eta^4, \qquad
\eta (t,x)=|\phi (t,x)|-\rho
\label{L2}
\end{equation}

\noindent
describes the {\it massive} gauge field and 1-component scalar field.
It is said that the
second component of Higgs field is "eaten" by the gauge field,
which gets the mass $e\rho$.

It should be noted that nonzero value of the gauge bosons masses
is a consequence of the spontaneously symmetry breaking
characterized by the perturbative vacuum expectation value $<\phi
>\ =\rho$. If the symmetry is unbroken ($\rho =0$) there are
massless bosons and 2-component scalar field.

As it was mentioned above all classical vacua (\ref{vac}) fall
into distinct classes, which are characterized by an integer
parameter $N$ (Chern-Simons number)~\cite{CDG}. Nielsen-Olesen
vortices~\cite{NO} play role of the instanton solutions in this
theory. Therefore different vacuum states $|\ N>$ can be connected
to each other by means of tunneling transitions. Thus the correct
vacuum is~\cite{JR,CDG}

\begin{equation}
|\ \Theta >\ \approx\sum_{N=-\infty}^{+\infty}e^{iN\Theta}|\ N>.
\label{theta}
\end{equation}

\noindent Nonzero value $<\phi >\ =\rho$ characterizes naive
perturbative vacuum with $N=0$ only. At the same time for the
correct vacuum $|\ \Theta >$ the scalar field expectation value is
zero~\cite{CDG}:

$$
<\Theta |\ \phi\ |\ \Theta >\ \approx
\sum_{N,M=-\infty}^{+\infty}e^{i(N-M)\Theta}<M|\ \phi\ |\ N>\ =
$$
\begin{equation}
=\sum_{N=-\infty}^{+\infty}<N|\ \phi\ |\ N>+\ O(e^{-S_{inst}})
\approx\rho\sum_{N=-\infty}^{+\infty}
e^{i{\textstyle \alpha}_N(x)}=0.
\label{VEV}
\end{equation}

\noindent where $S_{inst}\sim\rho$ is Euclidean instanton action,
$\alpha_{\scriptscriptstyle N}(x)$ is a set of the static
configurations which characterizes unequivalent vacua if
$A_0(t,x)=0$ and $e^{{\textstyle \alpha}_N(\pm\infty )}=1$.

The result (\ref{VEV}) means that {\it effective} potential for Higgs field
has only one minimum unlike classical potential.
This situation corresponds to the effective theory with zero mass
gauge field and 2-component scalar field.

\section{Screening of the confinement by
the unlike-charged particles production}
It is well known that confinement phenomenon exists in the 2-dimensional
Abelian Higgs model~\cite{CDG}.
Let one introduce two static external
charges by means of Wilson procedure~\cite{Wilson}.
We suppose that the interaction between external charges and gauge field
is described by the adding to the Lagrangian the term $j_{\mu}A_{\mu}$.
The energy
of the attraction between external charges is given by the following
formula (see for review~\cite{Rajar}):

\begin{equation}
E_{\Theta}(R)\sim e^{-S_{inst}}R\biggl[ cos\Theta -cos\biggl(\Theta+
\frac{2\pi q}{e}\biggr)\biggr],
\label{E}
\end{equation}

\noindent
where $R$ is a distance between two external particles with
the charges $+q$ and $-q$.

If $q=Ne$ the attraction disappears. It can be interpreted
as a screening of the external charges by means of the unlike-charged
scalar particle pairs production. However description of such charged
particles requires two-component field. Because of the total number
of the independent field components is conserved the theory possesses
{\it massless} gauge field only.
This statement contradicts with perturbative Higgs mechanism.
Thus possibility of the screening of the confinement
confirms the result of  the previous section.

\section{Conclusion}
As it is well known Higgs mechanism are used in Weinberg-Salam
theory for the $W^{\pm}$ and $Z^0$ bosons masses generating.
However there are not any experimental confirmation of Higgs
bosons existing. Let us suppose that our results can be
generalized into Weinberg-Salam theory. If instantons are not
gauge and lattice artifacts that standard model of electroweak
interactions contains massless gauge bosons and charged scalar
(Higgs) particles instead of the experimentally observed massive
gauge bosons. This argument leads to the necessity of the using of
an alternative mechanism of the $W^{\pm}$ and $Z^0$ bosons masses
generating without additional Higgs fields.

It should be noted, that in Weinberg-Salam theory instanton transitions
are strongly suppressed by the 't Hooft factor~\cite{Hooft}:

\begin{equation}
e^{-2S_{inst}}=e^{-\frac{4\pi }{\alpha_w}}\sim 10^{-160}.
\label{e}
\end{equation}

\noindent
However for the enough long time intervals the instanton
processes probability can increase

\begin{equation}
P \propto e^{-\frac{4\pi }
{\alpha_w}}\int dt_0dx_0
\end{equation}

\noindent
if we take into account the integration on the centre of the
instanton $t_0, x_0$ which is arbitrary space-time point.
For $t>e^{\frac{4\pi }{\alpha_w}}$
the probability of the instanton transitions is not negligible.
Thus instantons can restore gauge symmetry and violate Higgs mechanism.

It is interesting that instantons can dynamically break
the gauge symmetry and thus provide an alternative to the Higgs
mechanism~\cite{Dia}.

\end{sloppypar}

\end{document}